\documentstyle[aps,epsfig]{revtex}
\begin{document}
\def\vp{\varphi}
\def\e{\epsilon}
\def\b{\beta }
\def\i{\int (d\q)\,}
\def\tG{\tilde{G}}
\def\heta{\hat{\eta}}
\def\hc{\hat{c}}
\def\dheta{\hat{\eta}^\dag}
\def\dhc{\hat{c}^\dag}
\def\tsigma{\tilde{\sigma}}
\def\tg{\tilde{g}}
\def\tG{\tilde{G}}
\def\vPhi{\vec{\Phi}}
\def\vphi{\vec{\varphi}}
\def\vDelta{\vec{\Delta}}
\def\S{{\bf S}}
\def\r{{\bf r}}
\def\m{{\bf m}}
\def\a{{\alpha}}
\def\J{{\bf J}}
\def\k{{\bf k}}
\def\q{{\bf q}}
\def\l{\lambda}
\def\j{{\bf j}}
\def\n{{\bf n}}
\def\vn{{\bf v}}
\def\K{{\bf K}}
\def\M{{\bf M}}
\def\Q{{\bf Q}}
\def\E{{\bf E}}
\def\G{{\bf G}}
\def\B{{\bf B}}
\def\R{{\bf R}}
\def\A{{\bf A}}
\def\X{{\bf X}}
\def\F{{\bf F}}
\def\O{{\cal O}}
\def\T{{\cal T}}
\def\H{{\cal H}}
\def\L{{\cal L}}
\def\la{\langle}
\def\ra{\rangle}
\def\be{\begin{equation}}
\def\ee{\end{equation}}
\def\bea{\begin{eqnarray}}
\def\eea{\end{eqnarray}}
%\input epsf
% \draft command makes pacs numbers print
\draft
\title{Thermal vortex dynamics in a two-dimensional condensate}  

\author{R. \v{S}\'{a}\v{s}ik\cite{present},
Luis M. A. Bettencourt and Salman Habib}
\address{Theoretical Division,
Los Alamos National Laboratory, Los Alamos, New Mexico 87545}
\date{\today}
\maketitle

\begin{abstract}
We carry out an analytical and numerical study of the motion of an
isolated vortex in thermal equilibrium, the vortex being defined as
the point singularity of a complex scalar field $\psi(\r,t)$ obeying a
nonlinear stochastic Schr\"odinger equation. Because hydrodynamic
fluctuations are included in this description, the dynamical picture
of the vortex emerges as that of both a massive particle in contact
with a heat bath, and as a passive scalar advected to a background
random flow. We show that the vortex does not execute a simple random
walk and that the probability distribution of vortex flights has
non-Gaussian (exponential) tails.
\end{abstract}
% LAUR-99-3942

\section{Introduction}
Vortices and other topological field configurations play a fundamental
role in determining the properties of many fascinating materials and
control the physical mechanisms underlying several
applications. Examples include superfluids \cite{sonin},
superconductors \cite{brandt,farrell}, periodic solids \cite{mermin},
liquid crystals \cite{ss}, two-dimensional magnets \cite{msb},
propagating coherent light beams \cite{rss}, and perhaps even the hot
plasma that filled the very early Universe \cite{vs}. Common
aspects of the phenomenology of these diverse applications stem from
the mathematical similarity of the underlying field theories and their
classical static solutions.

Models of vortices as Brownian point particles, characterized by a
mass, mutual interactions, and damping, are commonly used to study the
behavior of superconductors and superfluids \cite{bmrefs}.  In this
picture the equation of motion for a single overdamped vortex, not
subject to any external forces, is extremely simple:
\begin{eqnarray}
\gamma \dot{\bf r}(t)&=&{\bf f}(t); \nonumber \\
\langle f_\mu(t)f_\nu(t') \rangle&=&\frac{2\gamma
k_BT}{M}\delta_{\mu,\nu}\delta(t-t'), \label{vbm}
\end{eqnarray}
where $\gamma$ is the damping coefficient, $M$ is the vortex mass,
${\bf f}(t)$ is a Gaussian thermal noise, $\mu,\nu\in\{x,y\}$, and
$\la\cdots\ra$ denotes canonical ensemble averaging.  It follows that
the vortex velocity distribution is also Gaussian.

This effective picture is extremely appealing---primarily because of
the drastic reduction of the number of degrees of freedom. However, to
our knowledge, it has not been shown to be derivable from the dynamics
of an underlying field theory. In this paper we study this question by
solving both analytically and numerically for the motion of the vortex
as an effective degree of freedom arising in a stochastic nonlinear
Schr\"odinger equation. It is well-known that the conventional
description of Brownian motion in terms of equations such as
(\ref{vbm}), that of a heavy particle interacting with light
scatterers, ignores the presence of hydrodynamic fluctuations. A more
complete description would allow us not only to test the validity of
the Brownian motion model for the vortex but also to compute
corrections to it and, possibly, to find new physical effects.

Our analytic approach rests on the use of a singular perturbation
expansion around a rigid vortex utilizing a low-temperature or
weak-noise expansion. We are able to derive a Fokker-Planck equation
for the single vortex distribution function which corresponds to that
of a passive scalar advected to a background flow [not just a simple
diffusion equation as would be predicted by Eqs.\ (\ref{vbm})]. In our
numerical work we are able to extract the diffusion constant for the
vortex motion which turns out to be in good agreement with the
theoretical prediction. Consistent with our theoretical analysis, the
vortex effective mass diverges logarithmically with the system
size. We also investigate the statistics of vortex flights and
demonstrate that the probability distribution function (PDF) has an
exponential tail implying nontrivial correlations in the background
thermal flow field.

In Section I we present the stochastic nonlinear Schr\"odinger
equation and discuss its physical relevance. In Section II we use the
low-temperature perturbative expansion about the vortex collective
coordinates to derive its equation of motion. Section III describes
our numerical techniques and the associated results. We end with
Section IV with a further discussion of our conclusions.

\section{The Stochastic Nonlinear Schr\"odinger Equation}

We study the motion of an isolated vortex by considering it as a
singularity of a classical stochastic nonlinear Schr\"odinger field
$\psi(\r,t)$ in two spatial dimensions. The equation of motion of this
model theory is $(a<0)$:
\begin{eqnarray}
(i\hbar-\gamma)\frac{\partial \psi}{\partial
t}&=&-\frac{\hbar^2}{2m}\nabla^2\psi 
+a\psi+b|\psi|^2\psi+\eta({\bf r},t); \nonumber\\
\langle\eta({\bf r},t)\eta^*({\bf
r}',t')\rangle&=&2k_BT\gamma\delta(t-t')\delta({\bf r} 
-{\bf r}'), \nonumber\\
\langle\eta({\bf r},t)\eta({\bf r}',t')\rangle&=&0.
\label{snls}
\end{eqnarray}
The damping coefficient $\gamma$ and the stochastic force $\eta(\r,t)$
model the coupling of the condensate to a heat bath, satisfying the
fluctuation-dissipation relation. Coefficients $a$, $b$ and $m$ are
assumed constant and independent of temperature; thermal effects are
fully described by the stochastic force.

The above equations (\ref{snls}) describe the phenomenology of
physically important systems such as atomic Bose condensates
\cite{stoof} and superfluid helium II \cite{p,gross}.  The same
equation (at $T=0$), amended as appropriate by the vector potential,
describes the order-parameter dynamics of type-II
superconductors \cite{schmid,gorkov}.  Although the above is known to
be strictly true only for a class of type-II superconductors with
magnetic impurities \cite{eliashberg}, it is often assumed that
Eqs.\ (\ref{snls}) apply to all type-II superconductors, unless
indicated otherwise.  We will henceforth study these equations as an
important paradigmatic field theory representing a broad range of
related physical phenomena.

It is convenient to make all quantities dimensionless by standard
substitutions.  We set $\psi = \tilde{\psi}\sqrt{-a/b}$, $\gamma =
\tilde{\gamma}\hbar$, $t=\tilde{t}\gamma/(-a)$, ${\bf r}=\tilde{\bf
r}\hbar /\sqrt{-2am}$, $\eta = \tilde{\eta}(-a)\sqrt{-a/b}$, and $k_B
T = -a\hbar^2/(2\tilde{\beta}mb)$.  Note that ${\bf r}$ is normalized
to a unit of length $\hbar/\sqrt{-2am}$, which is the Ginzburg-Landau
coherence length, usually denoted $\xi$.  With these substitutions,
and after dropping the tildes, Eqs.\ (\ref{snls}) become 
\bea 
(i/\gamma-1)\frac{\partial}{\partial t} \psi& =& -\nabla^2 \psi
+[|\psi|^2-1]\psi+\e\eta(\r,t); \nonumber\\ 
\langle \eta(\r,t) \eta^*(\r',t')\rangle &=& \frac{2}{\beta} 
\delta(t-t') \delta(\r-\r'), \nonumber\\ 
\langle \eta(\r,t) \eta(\r',t') \rangle &=& 0. 
\label{dsnls} 
\eea

The position of an isolated vortex, $\R(t)$, is defined via a contour
integral
\be 
\oint [\nabla\mbox{arg}\,\psi(\r,t)] \cdot d\,{\bf l}, \label{vortpos} 
\ee 
which equals $2\pi$ when the integration path encloses the point
$\R(t)$, and vanishes otherwise.

The Brownian motion of vortices as point particles has been argued to
hold primarily in the overdamped limit ($\gamma \rightarrow \infty$),
which is directly relevant to the superconducting case; this is also
the limit studied in this paper.  We envision a situation in which a
single vortex exists in the ground state of the field
$\psi(\r,t)$. 
In a thin superconducting film one can in
principle achieve the same effect by placing the sample in a magnetic
field of one flux quantum per sample area.  In the latter case there
are many other factors of practical importance, such as geometrical
and point pinning, which however will not be dealt with here.  The
vortex is assumed to be located near the center of the spatial extent
of the condensate, so that boundary effects can be neglected.  The
radius of the sample $\Lambda_c$ provides a natural infrared cutoff.
At sufficiently low temperature, thermally induced vortex-antivortex
pairs will be exponentially suppressed and we may assume that there is
only one vortex in a spatially bounded superfluid film at all
times. This assumption has been verified numerically for a range of
temperatures and sample sizes studied.

\section{Perturbative Analysis of Vortex Transport}

In this Section we utilize the singular perturbation expansion of Kaup
\cite{kaup} to extract the vortex position from the stochastic
equation of motion (\ref{dsnls}).  We first expand the field as 
\bea
\psi(\r,t) &=& \psi_0(\X)+\e\phi(\r,t); \label{psi}\\ 
\phi(\r,t) &=&\phi_1(\X,t) +\e \phi_2(\X,t)+ \cdots \label{phi}, 
\eea
where $\X\equiv (X,Y)$ is the comoving coordinate, measured relative
to a moving reference point; $\X(t) = \r-\R^0(t)$.  The absence of
explicit time dependence in $\psi_0(\X)$ in the comoving reference
frame indicates that it is the static vortex field in the absence of
fluctuations.  Tomboulis \cite{et} has shown that introduction of a
collective coordinate such as $\R^0(t)$, while conserving the total
number of dynamical degrees of freedom, is a canonical transformation.
The fluctuating field $\phi(\r,t)$ can be thought of as a
superposition of harmonic modes in the background of the rigid vortex
$\psi_0(\X)$. The phonons are gapless \cite{pit}, therefore they will
be an important consideration at any finite temperature.  The variable
$\e \ll 1$ plays the role of a small parameter as well as that of a
bookkeeping device to control the perturbation series \cite{gardiner}.
Another useful way to think about $\e$ emerges when one absorbs $\e$
into the definition of $\eta(\r,t)$ in Eq.\ (\ref{dsnls}): $\e$ then
reappears as the square root of temperature in the
fluctuation-dissipation relation in Eqs.\ (\ref{dsnls}). Therefore,
the small-$\e$ expansion is actually the {\it small temperature or
weak noise expansion.}  The equation of motion (\ref{dsnls}) is
required to hold at every order in $\e$, which results in a hierarchy
of equations for $\psi_0(\X)$, $\phi_1(\X,t)$, etc.

Because of the added fluctuations in Eq.\ (\ref{psi}), the actual
vortex position $\R(t)$ as the singularity of the full dynamical field
$\psi(\r,t)$ does not in general coincide with the coordinate
$\R^0(t)$, which by definition is the singularity of the rigidly
moving static vortex field $\psi_0(\X)$.  The two are related via
\be
\R(t)= \vec{\rho}\,[\R^0(t),\e\phi] =\R^0(t)+ \e\int d^2\r \left.  
\frac{\delta \vec{\rho}\,[\R^0(t),\phi]}{\delta \phi(\r)}
\right|_{\phi=0}\phi(\r,t)+\mbox{c.c.}+\cdots, \label{coord}
\ee
where $\vec{\rho}\,[\R^0(t),\e\phi]$ is a vector functional of the
fluctuation field $\e\phi(\r,t)$ and of the reference point $\R^0(t)$.
We define
\bea
\dot{\R}^0(t)&=&0 + \e{\bf v}^0_1(t) + \e^2{\bf v}^0_2(t) +\cdots,
\label{rdot1}
\\ 
\dot{\R}(t)&=&0 + \e{\bf v}_1(t) + \e^2{\bf v}_2(t) +\cdots,
\label{rdot2}
\eea
which makes explicit the fact that there is no vortex motion in the
absence of thermal fluctuations. Differentiating Eq.\ (\ref{coord})
with respect to $t$, we find at order $\e$,
\be
{\bf v}_1(t) = {\bf v}^0_1(t) + {\bf V}[\R(t),t], \label{v1}
\ee
where ${\bf V}[\R(t),t] \equiv \frac{d}{dt}\int
d^2\r\left. \frac{\delta\vec{\rho}\,[\R(t),\phi]}{\delta\phi(\r)}
\right|_{\phi=0}\phi(\r,t)+\mbox{c.c.}$ is a time-dependent velocity
field. In the last expression we replaced $\R^0(t)$ with $\R(t)$, as
the two are equal to $\O(\e)$. The statistical properties of ${\bf
V}[\R(t),t]$ are given by those of the excitation field
$\phi(\r,t)$\/, i.e., the phonons, and will therefore depend on
temperature.

Substituting Eqs.\ (\ref{psi}), (\ref{phi}), (\ref{rdot1}) and
(\ref{rdot2}) into Eqs.\ (\ref{dsnls}) and collecting powers of $\e$
we obtain, at order $\e^0$,
\be
0=-\nabla_\X^2\psi_0-[1-|\psi_0|^2]\psi_0. \label{lowest}
\ee
The properties of the static vortex solution are well known;
$\psi_0(\X) = f(|\X|)\exp[i\,\mbox{arg}(X+iY)]$, where $f(x)\sim x$
for $x\ll 1$ and $f(x) \sim 1$ for $x\gg 1$. At order $\e$ we find
\be
\frac{\partial}{\partial t}\left( \begin{array}{c}
\phi_1\\ \phi^*_1 \end{array} \right) =
\L \left( \begin{array}{c} \phi_1\\ \phi^*_1 \end{array}
\right)+{\bf v}^0_1\cdot\nabla_\X\left(
\begin{array}{c} \psi_0\\ \psi^*_0 \end{array} \right) -
\left( \begin{array}{c} \eta\\ \eta^* \end{array} \right),
\label{next}
\ee
where $\L$ is a Hermitean matrix 
\be
\L \equiv
\left( \begin{array}{lr}
\nabla_\X^2+[1-2|\psi_0|^2] & \psi_0^2\\
(\psi_0^*)^2 & \nabla_\X^2+[1-2|\psi_0|^2]
\end{array}
\right).
\ee
The noise $\eta(\X,t)$ remains white in space and time also in the
moving reference frame, just as in the original Eqs.\ (\ref{dsnls}).
Differentiating Eq.\ (\ref{lowest}) with respect to $X, Y$ we find
\be
\L\left( \begin{array}{c}
\partial_X \psi_0\\
\partial_X \psi^*_0
\end{array} \right)=\L\left(
\begin{array}{c}
\partial_Y \psi_0\\
\partial_Y \psi^*_0
\end{array} \right)=0.
\ee
Hence the null space of the linear operator $\L$ is spanned by the
eigenvectors
$(\partial_X \psi_0,\partial_X \psi^*_0)^T$ and 
$(\partial_Y \psi_0,\partial_Y \psi^*_0)^T$.
These generate uniform translations of $\psi_0$ in the $xy$ plane and
correspond to the Goldstone modes.

It turns out \cite{kaup} that all secular terms in the
$\e$\/-expansion vanish when we demand that the arbitrary perturbation
$(\phi,\phi^*)^T$ be orthogonal to the null space of the operator
$\L$.  With this condition in place [which also uniquely specifies the
reference point $\R^0(t)$] we now take the scalar product of
Eq.\ (\ref{next}) with the eigenvector $(\partial_X \phi_0, \partial_X
\phi^*_0)^T$, and integrate over the sample area to obtain
\be
0=\mbox{Re} \left\{ \int d^2\X\, (\partial_X\psi^*_0)[{\bf v}^0_1(t)
\cdot\nabla_\X\psi_0-\eta(\X,t)]
\right\}. \label{next2}
\ee

The same equation is obtained if we use the other eigenvector in the
scalar product, only now $\partial_X\psi^*_0$ is replaced by
$\partial_Y\psi^*_0$. Solving Eq.\ (\ref{next2}) for ${\bf v}^0_1(t)$
leads to
\be
{\bf v}^0_1(t)=\F(t), \label{stoch}
\ee
where the stochastic force $\F(t)$ is Gaussian with zero mean and
autocorrelator
\be
\la F_\mu(t)F_\nu(t')\ra = \frac{2}
{\beta M(\Lambda_c)}
\delta(t-t')\delta_{\mu\nu},
\ee
where $\mu,\nu \in \{X, Y\}$, and 
\be
M(\Lambda_c) \equiv \int^{\Lambda_c}d^2\X\,|\nabla_\X \psi_0(\X)|^2
\simeq 2\pi \ln\Lambda_c \label{mass}
\ee
can be interpreted as the (cutoff-dependent) {\it inertial mass} of
the vortex. The fully dimensional form of this equation is
$M(\Lambda_c)\simeq (2\pi\gamma^2/b)\ln(\Lambda_c/\xi)$.  This
quantity, in a different derivation, has received the same
interpretation by \v{S}im\'{a}nek \cite{simanek}.
The so-called vortex core mass\cite{suhl} 
is implicitly present in the first part of Eq.\ (18) and makes a constant
contribution to $M(\Lambda_c)$, whereas the asymptotic expression
in the latter part of Eq.\ (18) contains the
$\Lambda_c$\/-dependence which is dominant in the limit of large
$\Lambda_c$. 
In charged superfluids there is yet another contribution to the vortex
inertial mass, the so-called electromagnetic mass\cite{suhl}, which 
originates from the energy of the electric field generated by a moving
vortex. This contribution is absent in Eq.\ (18), which strictly applies
only to neutral superfluids.

Substituting for ${\bf v}^0_1(t)$ in Eq.\ (\ref{stoch}) from
Eq.\ (\ref{v1}) and using definitions (\ref{rdot1}) and (\ref{rdot2}),
we arrive at an equation of motion for the vortex, valid to $\O(\e)$,
\be
\frac{d}{dt}\R(t) =\e\F(t)+\e{\bf V}[\R(t),t].
\ee
As the last step we set $\e=1$ in this equation.

If we assume that ${\bf V}[\R(t),t]$ is slowly varying relative to the
noise ${\bf F}(t)$, then we may treat ${\bf V}[\R(t),t]$ as an
external field \cite{parisi} and arrive immediately at the
Fokker-Planck equation \cite{gardiner} for the density $c(\r,t) \equiv
\la \delta[\r-\R(t)]\ra$:
\be
\frac{\partial c(\r,t)}{\partial t} +\nabla\cdot[{\bf V}(\r,t)
c(\r,t)] = \sigma \nabla^2 c(\r,t), \label{fpe}
\ee
where the diffusion coefficient 
\be
\sigma \equiv \frac{1}{\beta M(\Lambda_c)}. \label{diffcon}
\ee
This equation is of the same form as that for a {\it passive scalar
advected to background fluid flow} ${\bf V}(\r,t)$ \cite{bj,pre}. It
is well established---particularly in the best-studied case of
incompressible fluid flow $\nabla\cdot{\bf V}(\r,t)=0$---that in
contrast to pure diffusion, an advected passive scalar may display
{\it non-Gaussian} (for example, exponential) tails in the probability
density function (PDF) $c(\r,t)$. These tails arise due to nontrivial
spacetime correlations in the advecting velocity field
\cite{bj,pre}. Interestingly, neither diffusion nor advection when
acting alone can produce non-Gaussian tails in the PDF's. In our case,
the non-Gaussian tails would appear in the PDF of vortex displacements
during a fixed time interval $\Delta t$ (vortex flights).  In order to
test this possibility without detailed knowledge of the velocity field
${\bf V}(\r,t)$, we study the thermal motion of an isolated vortex
numerically as described below.

\section{Numerical Analysis and Results}

In this section we describe our numerical methods used to solve
Eqs.\ (\ref{dsnls}) under the constraint that only a single vortex be
present in the simulation volume. An {\it unpaired} vortex is forced
into the sample by coupling the field $\psi(\r,t)$ to a static
external gauge potential $\A(\r)$, replacing $\nabla$ in Eq.\ (6) by
$\nabla + i\A(\r)$. If we choose $\A(\r) = -(2\pi/L^2)y\hat{x}$, there
will be on average one vortex per area $L\times L$ in an infinite 2D
sample. The introduction of $\A(\r)$ is a mathematical artifice and
does not represent a dynamical field. In the language of
superconductivity, $\A(\r)$ represents an external uniform magnetic
field that completely penetrates the sample without screening, and
vanishes in the thermodynamic limit $L\to\infty$.
This is a realistic approximation to a superconducting thin film geometry, 
as in this case the external magnetic flux must fully penetrate the
superconductor.
Screening is only manifested as an inhomogeneity of the flux density
threading the plane of the superconductor. The characteristic length scale
for these variations is\cite{dG}
$\Lambda=\lambda^2/d$, where $\lambda$ is the bulk magnetic
penetration length of the superconducting material,
and $d$ is the thickness of the film. We study the system in
the limit where $\Lambda$ is much larger than any other length of the
problem.  In this limit the flux density threading the superconductor
is uniform and screening can be neglected.
One might also 
worry that the derivation leading up to Eq.\ (18), which was done
explicitly for the case of the neutral superfluid, no longer applies
when $\A(\r)$ is introduced. While it is a fact that the
massless Goldstone boson in a superluid is in general replaced by a {\it 
massive} vector boson in a superconductor, the mass of this boson
vanishes in the thin film limit considered,
and the gapless phonon spectrum is restored, as
pointed out by \v{S}im\'{a}nek.\cite{simanek}
By the same token the 
electromagnetic vortex mass
vanishes in our thin film limit, as the moving vortex does
not generate any electric field.

We choose our system to be a square, $S=\la 0,L)\times\la 0,L)$.  We
wish to employ a variant of periodic boundary conditions so that the
entire $xy$ plane consists of physically equivalent copies of the
system at all times; by this we mean that the local superfluid density
$|\psi(\r,t)|^2$, and the supercurrent density
\be
{\bf J}(\r,t) =
\frac{i}{2}(\psi\nabla\psi^*-\psi^*\nabla\psi)+|\psi|^2\A 
\label{cond}
\ee
should both be periodic functions with periodicity $L$ in both spatial
directions.  To this end we impose quasi-periodic boundary conditions
on $\psi(\r,t)$ ($\r\in S$, $n_y$ and $n_y$ arbitrary integers):
\be
\psi(\r+n_xL\hat{x}+n_yL\hat{y},t)=\psi(\r,t) 
\exp[in_x\theta_x+in_y(2\pi x/L+\theta_y)].
\ee
The angles $\theta_x$ and $\theta_y$ are as yet arbitrary and may be
functions of time.  Upon discretization of Eqs.\ (\ref{dsnls}) we
observe that any given time dependence of $\theta_x$ and $\theta_y$
will in general affect the motion of $\psi(\r,t)$.  Yet $\theta_x$ and
$\theta_y$ are not dynamical degrees of freedom in the sense that
their motion is not determined by Eqs.\ (\ref{dsnls}). Therefore, we
are left with a problem of choosing the ``right'' time dependence for
$\theta_x$ and $\theta_y$. Fortunately the physical interpretation of
these variables is straightforward. In order to find it, we rewrite
the stochastic field equation in (\ref{dsnls}) in the form
\be
(i/\gamma-1) \frac{\partial}{\partial t} \psi= 
\frac{\delta H[\psi,\psi^*,\A]}{\delta \psi^*} +\e\eta(\r,t),
\ee
where 
\be
H[\psi,\psi^*,\A]\equiv \int
d^2\r\,\left[\frac{1}{2}|(\nabla+i\A)\psi|^2 
-|\psi|^2+\frac{1}{2}|\psi|^4\right]
\ee
is the Hamiltonian of the system (the condensate) disconnected from
the thermal reservoir.  One can show that
\be
\frac{\partial H}{\partial \theta_x}= 
\hat{x}\cdot\int_0^Ldy\,\left.\frac{\delta H}{\delta \A(\r)}\right|_{x=0}
\equiv \hat{x}\cdot\int_0^Ldy\,\left.{\bf J}_s(\r,t)\right|_{x=0}.
\ee
The first part of this equation is an algebraic equality; in the
second part we used the definition of the supercurrent, whose explicit
form appears in Eq.\ (\ref{cond}). Therefore, the first derivative of
the Hamiltonian with respect to $\theta_x$ is equal to the net
supercurrent passing through the edge of the sample perpendicular to
the $x$ axis. A completely analogous statement holds for the other
spatial direction as well.

If one is only interested in the thermal motion of the vortex in the
absence of external forces---such as, for example, the Magnus force
created by macroscopic superflow---the desired constraints on angles
$\theta_x$ and $\theta_y$ are
\be
\frac{\partial H}{\partial \theta_x}=\frac{\partial H}{\partial
\theta_y}=0. \label{zfe}
\ee  
These constraints define a {\it zero flow ensemble} \cite{dpa},
relevant for experiments with no externally imposed driving flow.  The
angles $\theta_x$ and $\theta_y$, though not dynamical variables, have
their values fixed by two dynamical constraints, (\ref{zfe}), at all
times.

As a result of the quasi-periodic boundary conditions, the isolated
vortex is part of an infinite (always perfectly square) vortex
lattice, with lattice constant $L$.  The physical cutoff $\Lambda_c$
needed to evaluate the vortex mass is estimated as $\Lambda_c \simeq
L/\sqrt{\pi}$ (this is when a square of side $L$ and a circle of
radius $\Lambda_c$ have equal areas).  Our numerical approach consists
of simultaneously solving the lattice version of Eqs.\ (\ref{dsnls})
on the spatial domain $S$, and the two dynamical constraints
(\ref{zfe}).  The numerical grid spacing $\delta$ must resolve both
the vortex core size $\xi$, which is the unit of length, and the
phonon correlation length $\xi_p$, which is the characteristic length
scale of the four-point correlator $\la
|\psi(\r,t)|^2|\psi(\r',t)|^2\ra$, and depends on temperature.  We
have found it adequate to set $\delta =0.15$ throughout the
temperature range studied.  Numerical stability of the second-order
stochastic Runge-Kutta method \cite{rk2} used to integrate
Eqs.\ (\ref{dsnls}) then allows a time step of $\tau=0.005$.  We study
sample sizes $L = 6$, 12, and 24, and temperatures $1/\beta = 0.0125$,
0.025, 0.05, and 0.1.

The vortex is clearly identifiable and no thermal vortex-antivortex
pairs are generated during the numerical runs. These ranged in
duration from $6\times 10^4$ to $6\times 10^5$, in dimensionless units
(up to $1.2\times 10^8$ time steps). The characteristic equilibration
time was $\sim 500$.  The vortex position is taken to be the
coordinate of the lower left-hand corner of the plaquette where it is
detected with the discretized version of the integral (\ref{vortpos}).

Fig.\ 1 shows $\langle [\Delta {\bf R}(t)]^2 \rangle$ for three
different $L$ and the same $\beta$.  It is apparent from the data
offsets that vortex motion is not purely diffusive. On the other hand,
the calculated diffusion coefficient $\sigma$ correctly approximates
the slopes of the numerical data at long times. Therefore we conclude
that the vortex mass, measured through $\sigma$, displays a
logarithmic divergence with the system size $L$.  The issue of the
value of the vortex inertial mass has been the subject of much
controversy in the literature \cite{nat,dl}.  Our results show that
even in the presence of random flow the logarithmic divergence of the
classical vortex mass persists. This stands in agreement with an
earlier result \cite{arovas} in the presence of coherent flow. It
would be interesting to investigate whether this divergence persists
in the presence of multiple vortices or in the limit of low damping.

An independent measure of diffusion is obtained through the concept of
the mean first passage time $\bar{t}(\ell)$ over the boundaries of the
square box of size $2\ell\times 2\ell$.  For pure diffusion one has
$\bar{t}(\ell) = 0.295\,\ell^2/\sigma$ \cite{sl}.  Fig.\ 2 reiterates
that the vortex motion is diffusive at long times. Deviations from
pure diffusion are manifest at short times that correspond to vortex
flights of $\O(1)$ and less.  The evidence from Fig.\ 2 also shows
that $\sigma$ depends linearly on temperature, as expected from
Eq.\ (\ref{diffcon}). This is a useful check on the quality of our
numerics---the vortex is not subject to spurious pinning due to the
discreteness of the underlying lattice. The flattening-out of the data
for large $\ell$ is due to finite sampling.

Finally in Fig.\ 3 we present the PDF of vortex flights. This
distribution has exponential tails (straight lines on a logarithmic
scale as in Fig. 3), in addition to the expected Gaussian profile
typical of Brownian motion. Our numerical results show that these
tails persist for all system sizes and $\beta$ studied, and the slope
decreases as the vortex mass $M(\Lambda_c)$ is increased (which
corresponds to a volume increase). The presence of exponential tails
has been linked to nontrivial time correlations in the advecting
velocity field \cite{bj}. The presence of finite correlation times can
be either a feature of a stochastic velocity field or can arise from
deterministic, chaotic motion (e.g., intermittency). In our case, we
expect the exponential tails to arise from the time correlations of
overdamped phonon modes which are allowed to exist in the fluctuation
spectrum of our theory. This aspect is presently under investigation.

\section{Discussion and Conclusions}

The particular models of vortex motion of the type described in the
Introduction typically ignore the presence of hydrodynamic modes.  In
contrast, real physical systems, in particular those described by
stochastic field theories, necessarily include these
effects. Consequently, as demonstrated here both analytically and
numerically, the motion of a single vortex must be viewed as diffusive
and randomly advected at the same time. The Fokker-Planck equation
(\ref{fpe}) makes explicit the formal analogy between single vortex
fluctuations and the dynamics of a passive scalar advected to random
fluid flow. The quantity $c(\r,t)$ is the ensemble-averaged
probability density of finding a vortex at $\r$ at an instant $t$.
The ensemble averaging is done over the fast degrees of freedom of the
heat bath. The slow, hydrodynamic, degrees of freedom of the
superfluid---the phonons---remain implicitly present in the velocity
field ${\bf V}(\r,t)$. This equation presents a picture of the vortex
as a diffusing massive particle subject to random advection. The
advection is apparently due to phonons, which propagate through the
vortex core randomly. As a result, the probability of large-scale
vortex flights is increased.

In turbulent fluid flow experiments with a dye as the passive scalar
it is possible to measure the dye concentration, analogous to our
$c(\r,t)$, directly, because individual dye particles do not interact
among themselves (apart from contact interaction).  Vortices, however,
interact logarithmically, so one cannot confirm the exponential tails
in $c(\r,t)$ directly by studying motion of a large aggregate of
vortices in a sample. Instead, one must study the motion of a single
vortex, and measure the statistics of vortex flights.

Examples where thermal vortex motion plays a key role include theories
of thermal depinning, flux creep, and dilute vortex lattice melting in
high-$T_c$ superconductors. Our current understanding of these
phenomena relies exclusively on the Brownian particle description.  It
would be interesting to re-examine these problems in the light of the
advected scalar aspect of vortex motion. An exciting new possibility
is a direct experimental observation of vortex motion in
two-dimensional high-$T_c$ superconductors by fast scanning tunneling
microscopy \cite{tak}. With this method the overwhelmingly more
frequent occurrence of large vortex flights in comparison with that of
a simple Brownian particle could be put to direct test.
  
\section{Acknowledgments}
We thank A.R. Bishop, D.P. Arovas, T. Hwa, and G. Lythe for useful
discussions.  Simulations were carried out at the Advanced Computing
Laboratory (ACL), Los Alamos National Laboratory, the Department of
Physics, Ohio State University, and at the National Energy Research
Scientific Computing Center (NERSC), Lawrence Berkeley National
Laboratory.

\newpage

\begin{figure}
\psfig{file=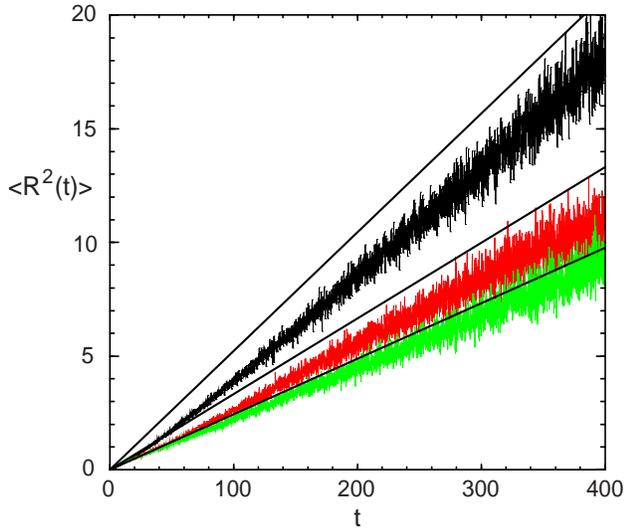,width=3.25in,height=2.8in}
\caption{The mean-squared vortex displacement vs.\ time, for three
different sample sizes (top to bottom): $L=6$, 12, and 24. The
temperature $1/\beta=0.1$ is the same for all plots. The straight
lines are the Einstein formula for pure diffusion, $\la
[\R(t)-\R(0)]^2\ra = 4\sigma t$, with $\sigma$ calculated from
Eq.\ (\ref{diffcon}).}
\end{figure}

\begin{figure}
\centerline{\psfig{file=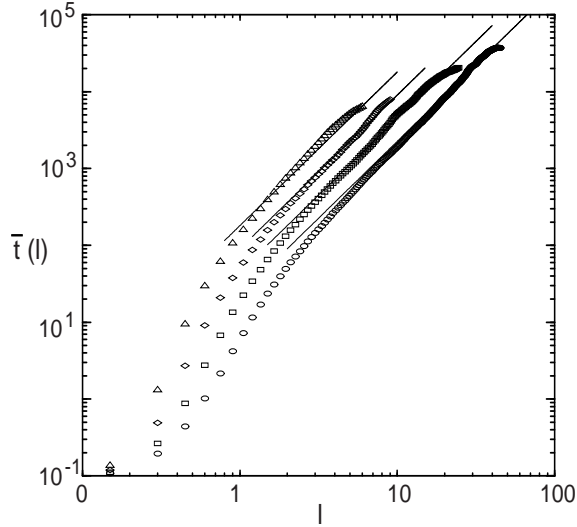,width=3.25in,height=2.8in}}
\caption{The mean first passage time vs.\ the box size, for four
temperatures (top to bottom): $1/\beta=0.0125$, 0.025, 0.05, and
0.1. System size $L=6$ is the same for all plots. Straight lines are
the theoretical predictions for pure diffusion, with $\sigma$
calculated from Eq.\ (\ref{diffcon}).}
\label{fig2}
\end{figure}

\begin{figure}
\centerline{\psfig{file=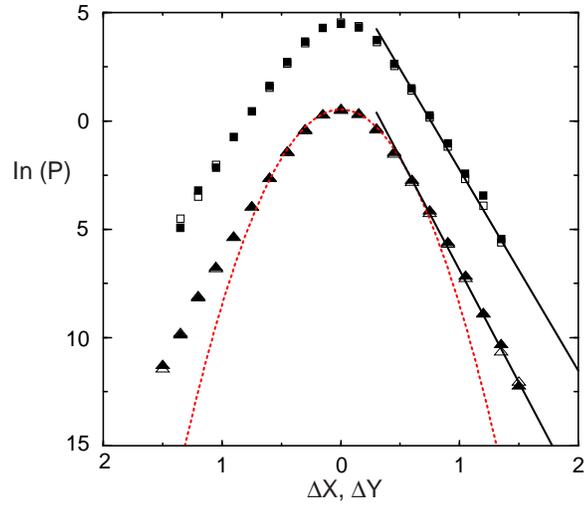,width=3.25in,height=2.8in}}
\caption{The PDF of vortex flights, for $L=6$ (triangles) and $L=12$
(squares, shifted by four orders for clarity).  $1/\beta=0.1$, and
time interval between measurements $\Delta t= 0.1$.  Full and empty
symbols correspond to ${\cal P}(\Delta x)$ and ${\cal P}(\Delta
y)$. The dotted line is the Gaussian distribution with the same rms
deviation as the numerical data for $L=6$.  The straight lines are
drawn with slopes corresponding to numerical data.}
\label{fig3}
\end{figure}

% figures follow here
%
% Here is an example of the general form of a figure:
% Fill in the caption in the braces of the \caption{} command. Put the label
% that you will use with \ref{} command in the braces of the \label{} command.
%
% \begin{figure}
% \caption{}
% \label{}
% \end{figure}

% tables follow here
%
% Here is an example of the general form of a table:
% Fill in the caption in the braces of the \caption{} command. Put the label
% that you will use with \ref{} command in the braces of the \label{} command.
% Insert the column specifiers (l, r, c, d, etc.) in the empty braces of the
% \begin{tabular}{} command.
%
% \begin{table}
% \caption{}
% \label{}
% \begin{tabular}{}
% \end{tabular}
% \end{table}

\end{document}